\documentclass[runningheads]{llncs}
\usepackage[T1]{fontenc}
\usepackage{graphicx}
\usepackage{hyperref}
\usepackage{xcolor}

\usepackage[numbers]{natbib}
\usepackage{enumitem}
\usepackage{amsmath, amssymb, amsfonts}
\usepackage{float}
\usepackage{cleveref}
\usepackage{subcaption}
\usepackage{listings}
\usepackage{booktabs}

\newcounter{mainrq} % Counter for main RQs
\newcounter{subrq}[mainrq] % Counter for sub-RQs, resets with every main RQ

\renewcommand{\themainrq}{\arabic{mainrq}}
\renewcommand{\thesubrq}{\themainrq.\arabic{subrq}}

\crefname{mainrq}{RQ}{RQs}
\crefname{subrq}{Sub-RQ}{Sub-RQs}

\newlist{subquestions}{enumerate}{1}
\setlist[subquestions,1]{
    label=\textbf{Sub-RQ\themainrq.\arabic*:}, % How the label looks in the list
    ref=\thesubrq,                             % How the reference number looks
    leftmargin=*,
    before=\let\olditem\item\renewcommand{\item}{\refstepcounter{subrq}\olditem}
}

\usepackage{setspace} % Needed for \setstretch

\lstdefinestyle{mypromptstyle}{
  backgroundcolor=\color{lightgray!15},
  basicstyle=\ttfamily\footnotesize\setstretch{0.8},
  breaklines=true,
  frame=single,
  framerule=0pt,
  framesep=2pt,
  rulecolor=\color{black},
  showstringspaces=false,
  tabsize=2,
  captionpos=b,
  aboveskip=0.5em,
  belowskip=0.5em,
  lineskip=0pt                   % Reduce vertical space between lines
}
\newcommand{\headernodot}[1]{\vspace{0.8mm}\noindent\textbf{#1}}
\newcommand{\header}[1]{\headernodot{#1.}}

\newcommand{\TODO}[1]{|\textcolor{red}}

\begin{document}
\title{How Do LLMs Cite?}
% \subtitle{A Mechanistic Interpretation of Attribution in Retrieval-Augmented Generation}
\subtitle{A Mechanistic Interpretation of Attribution in RAG}
\author{Ian van Dort\inst{1}\orcidID{0009-0008-5881-760X} \and
Maria Heuss\inst{1}\orcidID{0000-0001-5360-9627}}
% \author{Anonymous}
%    \institute{Anonymous Institution} 
%
\authorrunning{I. van Dort and M. Heuss}
% \authorrunning{Anonymous}
\institute{University of Amsterdam\\
\email{ian.van.dort@student.uva.nl, m.c.heuss@uva.nl}}
\maketitle              % typeset the header of the contribution
\begin{abstract}

With the rising popularity of large language models (LLMs) for various applications ranging from every-day information seeking to high stakes applications such as complex medical question answering, hallucinated content poses a serious risk. % to the trustworthiness of such LLM-based information retrieval (IR) systems.
Retrieval-Augmented Generation (RAG) aims to enhance the trustworthiness of LLMs by grounding their outputs in external documents, often using inline citations for verifiability. 
However if those citations are not faithful, i.e. if they do not accurately reflect the source of the information during the answer generation, user trust might be misguided by the sheer existence of such references to trusted sources. 
% With this work we take a step towards understanding the mechanisms that underlie the generation of inline citations and their faithfulness to the source documents. 
We argue that to understand citation faithfulness and to develop a reliable framework for the evaluation of citation faithfulness, a mechanistic approach that considers the model internals, rather than mere observations on the input/output of the model, is necessary. 
% the faithfulness of these citations---whether the model genuinely uses a source to generate an answer---remains a critical, unverified assumption. 
This paper offers the first mechanistic account of how a large language model decides whether to attach an inline citation while answering a factoid question. 
%One sentence on findings.
% Using the Llama-3.1-8B-Instruct model in a controlled experimental environment based on the PopQA dataset, we employ an activation patching approach. 
% We map the underlying mechanism responsible for citation, discovering that it is not a single, localized component but a distributed, multi-stage ``attributional ensemble'' of attention heads and MLP layers.
% We show that amplifying or attenuating only those critical heads and MLPs repairs over 90\% of missed citations and eliminates 69\% of spurious ones on PopQA without harming answer accuracy. 
% Although gains on the multi-document HotpotQA benchmark are modest, the same component set still moves citation rates in the intended direction, indicating that the underlying mechanism is not dataset-specific. 
Through activation patching we identify an ``attributional ensemble'' of attention heads and MLP layers that are responsible for the citation generation process in Llama-3.1-8B-Instruct. 
% We reveal a potential disconnect between the model's apparent reasoning and its internal computational pathway, suggesting that inline citations can create a false sense of security. 
Our findings suggest that citation decisions rely heavily on shallow heuristics such as entity co-reference matching, raising concerns about the trustworthiness of such citations.
%, raising questions about whether inline citations in current RAG systems reliably indicate that retrieved documents causally influenced answer generation.

% \maria{@Ian, not sure if this is really the right way to formulate what we are showing in this last sentence. Can you see whether you find a formulation that is more on point and not overpromising?}

% By uncovering a fallible, heuristic-based mechanism for a behavior central to AI usefulness, this work provides a critical foundation for developing more genuinely trustworthy and architecturally verifiable RAG systems.

\end{abstract}


\section{Introduction}

Large Language Models (LLMs) are increasingly deployed in high-stakes domains, from medical question answering~\citep{yang2023large} to legal research~\citep{seabrooke2024survey} and educational tools~\citep{kasneci2023chatgpt}. These systems promise to democratize access to specialized knowledge that was previously available only to experts or those with substantial resources. However, LLMs are prone to ``hallucinations'': plausible but factually incorrect outputs~\citep{ahmad_creating_2023}. When a lawyer relies on a fabricated case citation~\citep{merken_new_2023} or a clinician acts on incorrect medical information~\citep{modi_assessing_2025}, the consequences can be severe.

Retrieval-Augmented Generation (RAG) addresses this by grounding model outputs in external documents and providing citations. The appeal is straightforward: users can verify claims against sources. Yet this verification promise matters most for communities that rely on AI as a primary information source but lack expertise to independently assess the underlying documents. If citations are merely cosmetic---appended after the fact rather than genuinely driving the answer---then RAG systems risk creating a false sense of security that could deepen information disparities rather than close them~\citep{liu_evaluating_2023}.

The problem is that citation correctness (does the source factually support the claim?) does not guarantee citation faithfulness (did the model actually use the source to generate the claim?). Consider a medical RAG system asked about drug interactions. If the model recalls the answer from training data and then attaches a citation to a document that happens to confirm it, the citation is correct but unfaithful. Worse, research shows that explanations, even misleading ones, increase user trust in AI outputs~\citep{sadeghi2024explaining}. In dense, technical domains where users cannot easily verify sources themselves~\citep{magesh_hallucination-free_2024, lee_beyond_2006}, unfaithful citations lend misinformation an air of credibility.

\header{Citation Faithfulness as a Property of Trustworthy Attribution}
We define citation faithfulness as a citation occurring only when the model's internal computation genuinely used the cited document to construct the answer~\citep{wallat_correctness_2024}. The challenge is that faithfulness is not observable from input-output behavior alone. A model can produce identical text whether it retrieved information from a document or hallucinated from memory. To assess faithfulness, we must look inside the model.

\header{Towards Understanding Citations with Mechanistic Interpretability}
This paper uses mechanistic interpretability to reverse-engineer how Llama-3.1-8B-Instruct decides whether to emit an inline citation in a RAG context. Instead of treating the model as a black box, we use activation patching to identify which internal components (specific attention heads and MLP layers) influence citation behavior. We then test whether these components can be manipulated: inducing citations and suppressing spurious ones.

Why does this matter? If citations are generated by shallow heuristics (e.g., matching entity names between document and question) rather than through genuine engagement with document content during answer construction, then current RAG systems offer only an illusion of verifiability. Understanding the mechanism is the first step toward building evaluation frameworks that can distinguish faithful from unfaithful attribution---critical for deploying AI systems that communities with limited technical resources can actually trust.

\noindent The main contributions of this work are:

\begin{enumerate}
    \item Mapping the distributed, multi-stage ``attributional ensemble'', a collection of specific attention heads and MLP layers, that collaboratively implements the citation mechanism.
    \item Demonstrating the role of this ensemble by performing targeted interventions that successfully correct citation failures, such as inducing missing citations or suppressing spurious ones.
    \item Validating the generalizability of these findings by applying the same interventions in a more complex, multi-document question-answering setting and achieving a significant improvement in citation accuracy.
\end{enumerate}

\section{Related Work}\label{chapter:rw}
Large language models (LLMs) often ``hallucinate'', producing factually incorrect information~\citep{ahmad_creating_2023}.  To attempt to overcome these issues, RAG helps LLMs by retrieving relevant documents from external knowledge bases, and using those documents to draft better outputs~\citep{tonmoy_comprehensive_2024}. It does this by coupling a retriever, which identifies relevant documents or data, with a generator, typically an LLM, to produce outputs grounded in external knowledge, often providing inline citations to enhance verifiability and user trust~\citep{lewis_retrieval-augmented_2020}. This approach can enhance the accuracy and credibility of the generated output, especially for tasks which require extensive up-to-date or domain-specific knowledge~\citep{gao_retrieval-augmented_2024}. However, the presence of a citation does not guarantee trustworthiness. A critical, unverified assumption is that these citations are faithful: that is, the cited document had a genuine causal influence on the generation of the corresponding statement~\citep{wallat_correctness_2024, jacovi_towards_2020}. A model might generate an answer from its parametric memory and append a citation that is merely plausible or correct, but not causally responsible, creating a misleading facade of credibility. This distinction is vital in high-stakes domains like law and medicine, where unfaithful citations could lead to misplaced trust in incorrect information~\citep{magesh_hallucination-free_2024,lee_beyond_2006}. In this work we build on the definition of citation faithfulness from~\citep{wallat_correctness_2024}, but take a step towards approaching the problem by understanding how the model generates citations. 

\subsection{Approaches to Evaluating and Performing Attribution}
Research on RAG attribution has primarily followed three paths.

\subsubsection{External, Output-Based Methods:} The most common approach uses external models to evaluate the relationship between the generated text and the source after generation is complete. These methods often frame the task as Natural Language Inference (NLI), checking if the source text (premise) entails the generated claim (hypothesis)~\citep{bohnet_attributed_2022,muller_evaluating_2023}. While useful for measuring citation correctness, these post-hoc methods cannot assess faithfulness, as they have no access to the model's internal generation process. They answer ``Could this source support the claim?'' not ``Did this source cause the claim?''.

\subsubsection{Generation-Integrated Self-Citation:}
To streamline attribution, models can be prompted or fine-tuned to generate citations inline with the answer~\citep{gao_enabling_2023}. This approach is used by many state-of-the-art RAG systems and is the focus of our investigation. While this integrates attribution into the generation process, it does not guarantee faithfulness. As prior work on self-explanations has shown, a model's stated rationale for its output does not necessarily align with its true computational process~\citep{madsen_are_2024, turpin_language_2023}.

\subsubsection{Mechanistic and Perturbation-Based Methods:}
A smaller body of work has sought to establish a more causal link between source and output by analyzing model internals. Techniques like gradient-based saliency~\citep{qi_model_2024}, Jensen-Shannon Divergence~\citep{li_attributing_2025}, hidden-state analysis~\citep{phukan_peering_2024} and perturbation analyses~\citep{cohen-wang_contextcite_2024} identify which parts of the context most influence the final output. These methods offer a stronger claim to faithfulness but are often computationally expensive, require full model access, rely on coarse sentence-level ablations that can miss nuanced influences, and have not yet been used to map the full, end-to-end mechanism of how a citation decision is made.

% While prior work has used external methods to validate RAG outputs or developed coarse-grained causal metrics, the internal computational mechanism governing how an LLM decides to produce a citation remains a black box. This paper provides the first mechanistic account of this process. We leverage techniques from mechanistic interpretability—specifically probing to identify information representations and interventions (signal injection and activation patching) to test their function and map their circuitry ~\citep{meng_locating_2022, alain_understanding_2018}. By moving beyond input-output correlations, we reverse-engineer the ``attributional ensemble'' responsible for inline citation, revealing the heuristic-based, multi-stage algorithm the model has learned for this critical behavior.

\section{Methodology}\label{chapter:method}
We designed a minimal, controlled experimental pipeline to study when and why a RAG-style model emits inline citations. To avoid confounds from a full retrieval stack, each example is run twice with the same prompt: once with a relevant supporting document and once with an irrelevant (distractor) document. Comparing these two conditions isolates behaviour driven by the supplied context.

Below we summarise the dataset, model, prompt selection, and the component-level evaluation used to locate and validate mechanisms that correlate with citation behaviour.

\subsection{Dataset and model}\label{methodology:section:dataset_model}
We base our experiments on a citation-focused adaptation of PopQA~\citep{mallen_when_2023}. PopQA's short, single-fact questions and clear answers make automated correctness checks straightforward. For each question we generate (i) a short supporting document that contains the answer and (ii) a distractor document.
% Distractors were created either by random sampling of other factual sentences (length-matched) or by randomly sampling structure-matched factual sentences that mirror subject/object token lengths and relation type. The latter preserves token-level structure, enabling precise component-level comparisons.
Distractors were created by randomly sampling structure-matched factual sentences that mirror subject/object token lengths and relation type. This preserves token-level structure, enabling precise component-level comparisons.

All experiments use the Llama-3.1-8B-Instruct model, chosen because it reliably follows citation-style instructions and is supported by TransformerLens~\citep{grattafiori_llama_2024,nanda_transformerlens_2022}.
% Implementation details and dataset generation code are provided in the Experimental Setup (\S\ref{sec:exp-model}, \S\ref{sec:exp-dataset-popqa}).

\subsection{Prompt design}\label{sec:prompt-design}
Prompt wording affects both answer correctness and citation propensity. We therefore selected a short, zero-shot instruction that (a) asks the model to add an inline citation marker (e.g. "[1]") only when a claim is supported by the provided document, and (b) asks the model to answer from prior knowledge without a citation when the document is irrelevant. An example of the final prompt structure we use is:
\begin{lstlisting}
<|begin_of_text|><|start_header_id|>system<|end_header_id|>

Cutting Knowledge Date: December 2023
Today Date: 26 Jul 2024

When answering the question: 1) If the document contains relevant information, use it and cite it with [1]. 2) If the document isn't relevant, answer based on your knowledge without citation. 3) Never cite information you didn't get from the document. 4) Citations should appear immediately after the specific information from the document.<|eot_id|><|start_header_id|>user<|end_header_id|>

Document 1: Kampala is the capital of Uganda. 

Question: What is the capital of Uganda?<|eot_id|><|start_header_id|>assistant<|end_header_id|>

The capital of Uganda is
\end{lstlisting}\label{prompt_structure}

% Section~\ref{sec:exp-prompt-sel} describes the prompt search and selection procedure.

\subsection{Component-level evaluation via activation patching}\label{sec:methodology-activation-patching}
To locate internal components associated with citation behaviour we perform activation patching between the relevant and distractor runs. Activation patching replaces selected intermediate activations in one run with activations from the other and measures the resulting change in the model's preference for starting a citation. We operationalise this preference as the logit difference between the citation-initiating token (the token for the character " [") and the sentence-final token ".".

To reduce variance, the analysis is restricted to a homogeneous subset of examples that share a single structural template and where the clean (relevant) and corrupted (distractor) runs produce clear, opposite top-token preferences at the final step. We report a normalized logit-difference score that measures the fraction of behavioural change recovered (or induced) by a patch; the formula is given in Equation~\ref{equation:normalized-logit-diff}.

\begin{equation}\label{equation:normalized-logit-diff}
    \text{Score} = \frac{\text{Patched LogitDiff} - \text{Corrupted LogitDiff}}{\text{Clean LogitDiff} - \text{Corrupted LogitDiff}}
\end{equation}

We systematically test three locations: full residual replacement, attention output, and MLP output.
% For attention we further decompose heads into query/key/value outputs and attention patterns to localise effects more precisely. 
Token-position-level patching is limited to the portion of the prompt where the two runs diverge to avoid redundant edits of identical context.

\subsection{Evaluation of identified components}\label{methodology:evaluation}
We select important components based on the average normalized logit difference score for that component when patching. Identified components are validated in three ways: targeted interventions on failure cases to increase or decrease a component's contribution (missed or spurious citations); a unified intervention applied across the full PopQA set to test generalisation; and testing the same components in a multi-document QA task (HotpotQA) where the citation metric demands the correct set of supporting documents~\citep{yang_hotpotqa_2018}. The code for dataset creation and all experiments is available at: \url{https://github.com/ivdort/thesis-RAG-MI}.

\section{Results and Discussion}\label{sec:evaluation}
We present results in four steps. First, we analyze residual-stream patching in denoising and noising settings to localize influential layers and token positions. Second, we decompose effects by component (attention vs. MLP).
% followed by head-level and sub-component analyses to refine localization
Third, we validate and steer the identified components via targeted interventions on PopQA. Finally, we test generalization to the multi-document HotpotQA setting. Experimental details for patching are given in Section~\ref{sec:methodology-activation-patching}, and the evaluation protocol is in Section~\ref{methodology:evaluation}. Effects throughout are quantified using the normalized logit-difference score (Equation~\ref{equation:normalized-logit-diff}).

For clarity, we show denoising where it highlights components that positively contribute to citation (clean into corrupted), and we show noising where it highlights components that are necessary (corrupted into clean). In practice both modalities tell a consistent story, though noising typically produces larger-magnitude effects due to the mechanism’s conjunctive dependencies.

\begin{figure}[h]
    \centering
    \includegraphics[width=0.9\linewidth,trim=0 0.7cm 0 5cm,clip]{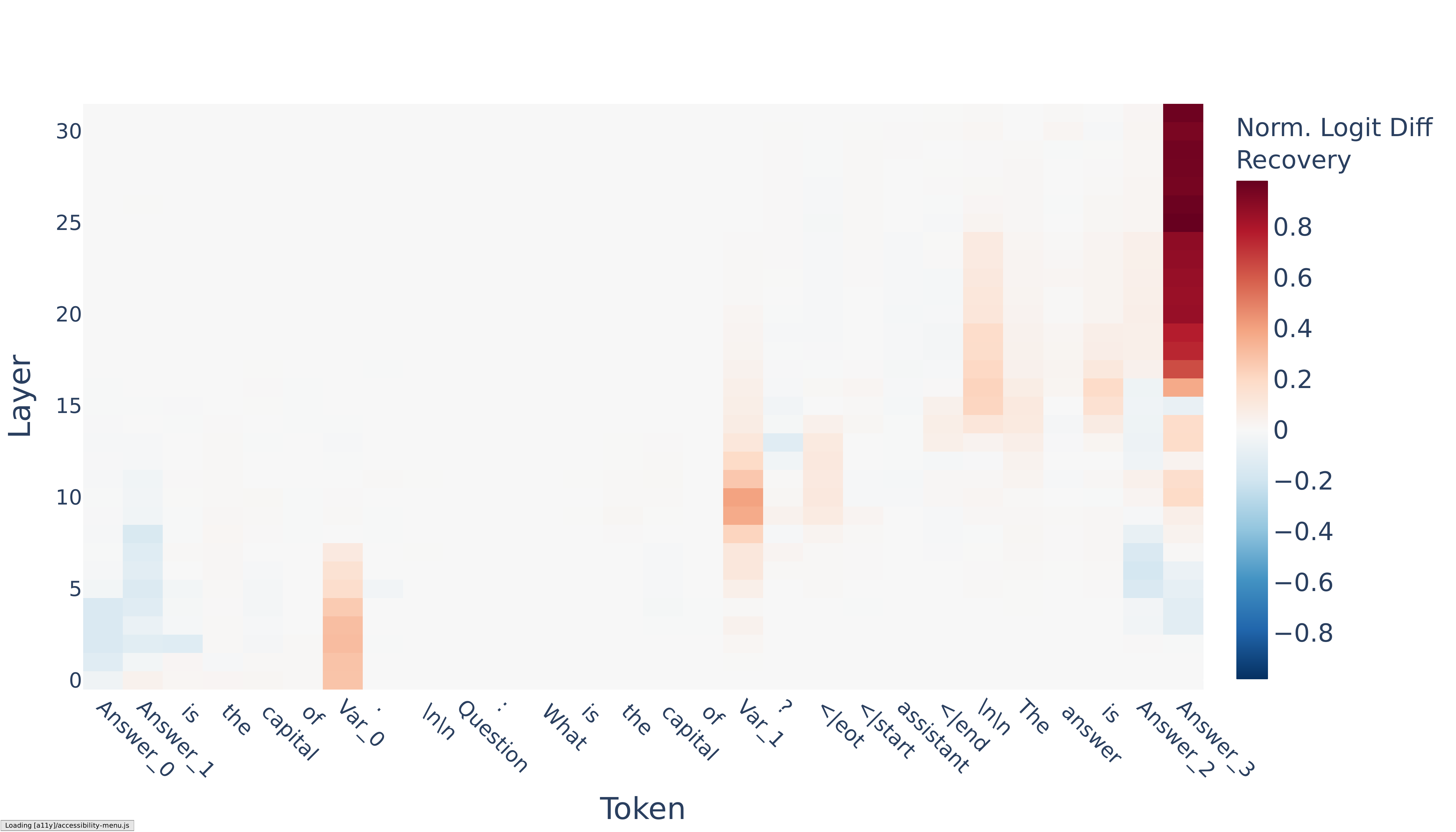}    \caption{Residual stream (denoising).}
    \label{fig:residual_denoising}
\end{figure}

\begin{figure}[]
	\centering
	\begin{subfigure}[b]{0.505\textwidth}
		\centering
		\includegraphics[width=\linewidth,trim=1.5cm 0.9cm 0 5cm,clip]{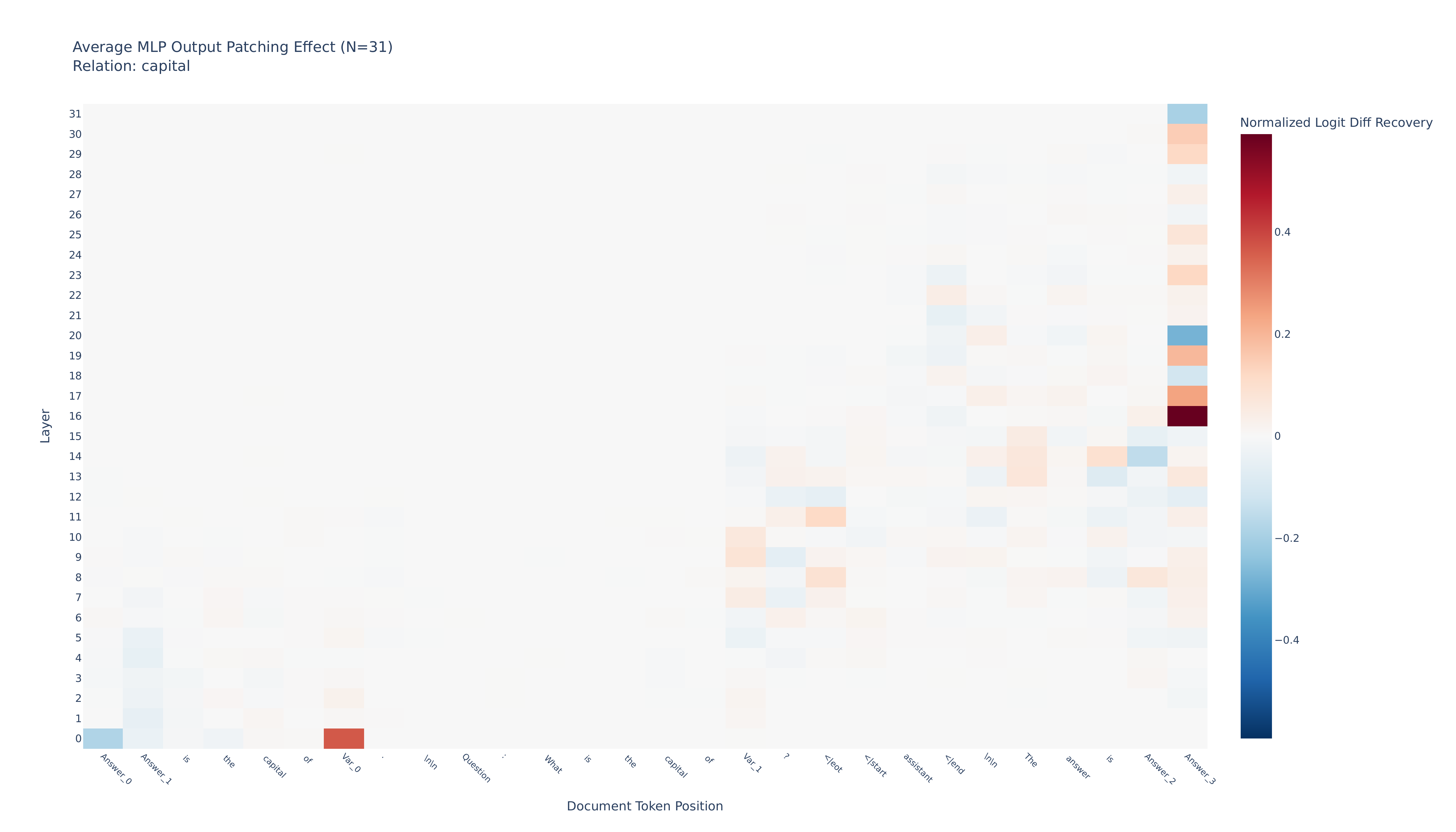}
		\caption{MLP output denoising.}
	\end{subfigure}\hfill
	\begin{subfigure}[b]{0.48\textwidth}
		\centering
		\includegraphics[width=\linewidth,trim=0 0.7cm 0 5cm,clip]{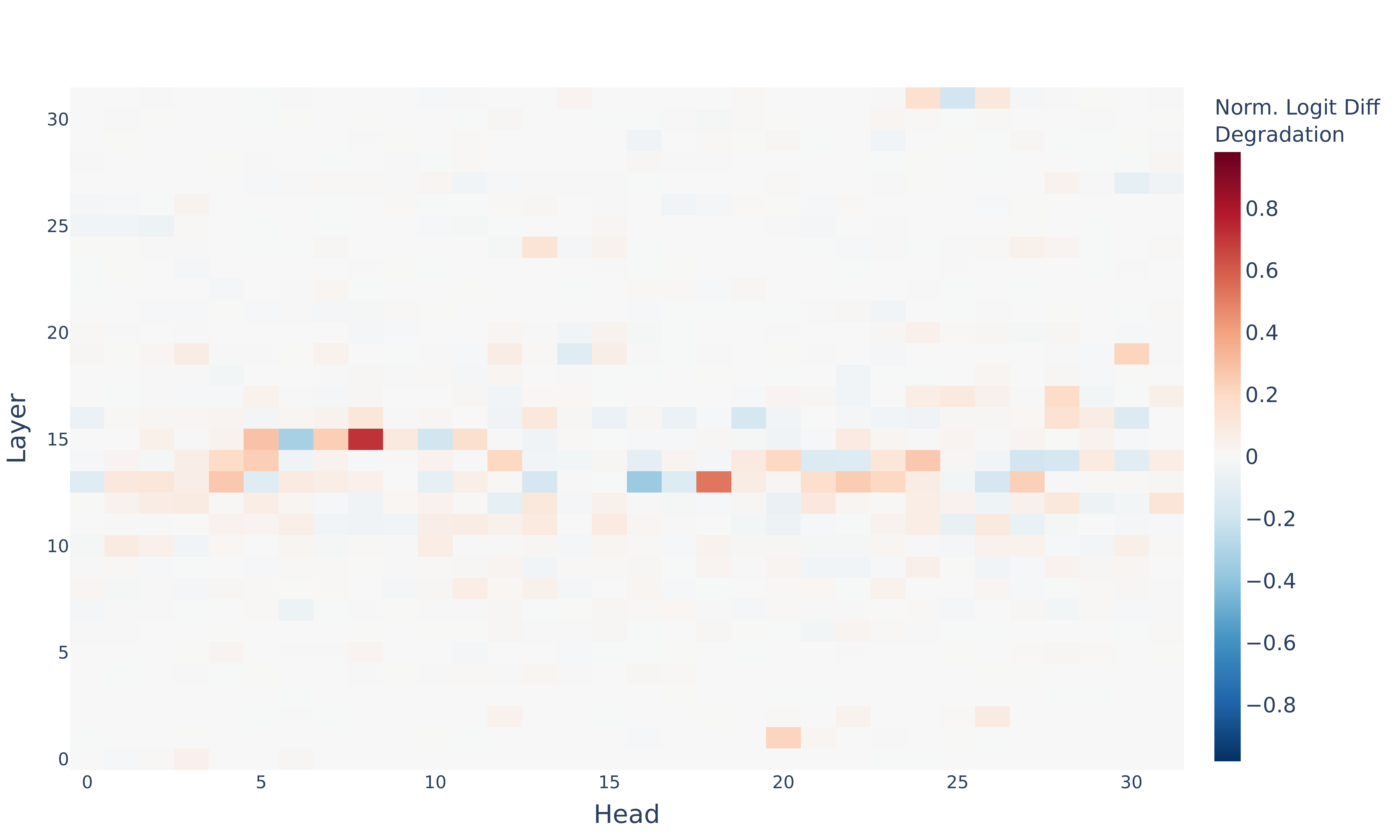}
		\caption{Attention head noising.}
	\end{subfigure}
	\caption{Key mechanistic signals: (a) token/layer regions whose clean MLP activations restore citation; (b) individual attention heads whose corrupted outputs most strongly disrupt citation (e.g., L15H8).}
	\label{fig:mechanism_core}
\end{figure}

\paragraph{Core Structural Pattern: A Distributed Attributional Ensemble.} 
% In response to RQ~\ref{rq:2.2}, 
We find that the decision to cite is governed not by a single ``citation head'' but by a distributed and fragile \textbf{attributional ensemble} of many attention heads and MLPs, where ``fragile'' means that small corruptions at many sites strongly reduce citation, while single clean patches rarely fully induce it. As shown in Figures~\ref{fig:residual_denoising} and~\ref{fig:mechanism_core}, these components' contributions accumulate across model depth in a process that is easily disrupted (via noising) but difficult to induce (via denoising), consistent with an AND-like dependency structure where multiple conditions must be met.

\paragraph{Early Layers: Entity Enrichment and Duplication Matching.}
The model first identifies a correspondence between the document and the question using a multi-step heuristic. An early MLP (layer 0) performs an ``entity enrichment'' step on the queried entity token where it first appears in the document---for example, on `Uganda' in ``Document 1: Kampala is the capital of \textbf{Uganda}.'' (this token is Var\_0 in Figure~\ref{fig:mechanism_core}a). This enriched representation is then passed to the second occurrence of the same entity in the question, ``What is the capital of \textbf{Uganda}?'' (Var\_1). The significance of this pathway is striking: while the text of the question is identical in both clean (relevant document) and corrupted (irrelevant document) runs, patching the internal activation at Var\_1 is sufficient to restore or break the citation behavior. This suggests the activation carries an internal flag for cross-context identity, effectively ``marking'' the entity as grounded in the provided document. This mechanism, similar to co-reference heuristics found in prior work~\citep{wang_interpretability_2022}, creates a durable, enriched representation that is then propagated forward to influence the final citation decision, as seen in the information path in Figure~\ref{fig:residual_denoising}.

\paragraph{Mid-Layers: Processing Hub (Layers ~10--20).} A dense band of attention heads and MLPs in middle layers contribute moderate, distributed gains (denoising) or large degradations (noising). Certain heads (e.g., a representative necessary head such as L15H8, highlighted in Figure~\ref{fig:mechanism_core}b) appear individually critical: corrupting them removes >60\% of citation behavior. Yet no single head is sufficient to induce high recovery alone (typical positive denoising contributions 10--30\%). We hypothesize that these middle-layer components implement conjunctive checks whose effects accumulate, but we note that activation patching localizes nodes rather than directed edges; validating pathways will require path patching.

\paragraph{Late Decision Assembly.} The final answer token position aggregates upstream signals, visible as a cluster of effects at the final answer tokens in Figures~\ref{fig:residual_denoising} and~\ref{fig:mechanism_core}a. Late-layer MLPs (e.g., layer 16) and selective attention heads (including a decisive late head such as L24H13) increase the logit of the citation marker over a competing termination token (`.'). Due to the logit-difference metric's binary framing, we cannot always disambiguate whether components promote citation or suppress sentence termination, an inherent interpretive limitation.

% \paragraph{Query vs. Value Roles.} Sub-component patching of key heads showed query vectors dominate attention \emph{pattern} formation (where to look), while value vectors supply decisive content written into the residual stream. Corrupting queries often reroutes attention sufficiently to break the downstream chain, whereas clean value vectors more consistently provide additive recovery.

\begin{figure}[h]
    \centering
    \includegraphics[width=0.86\linewidth]{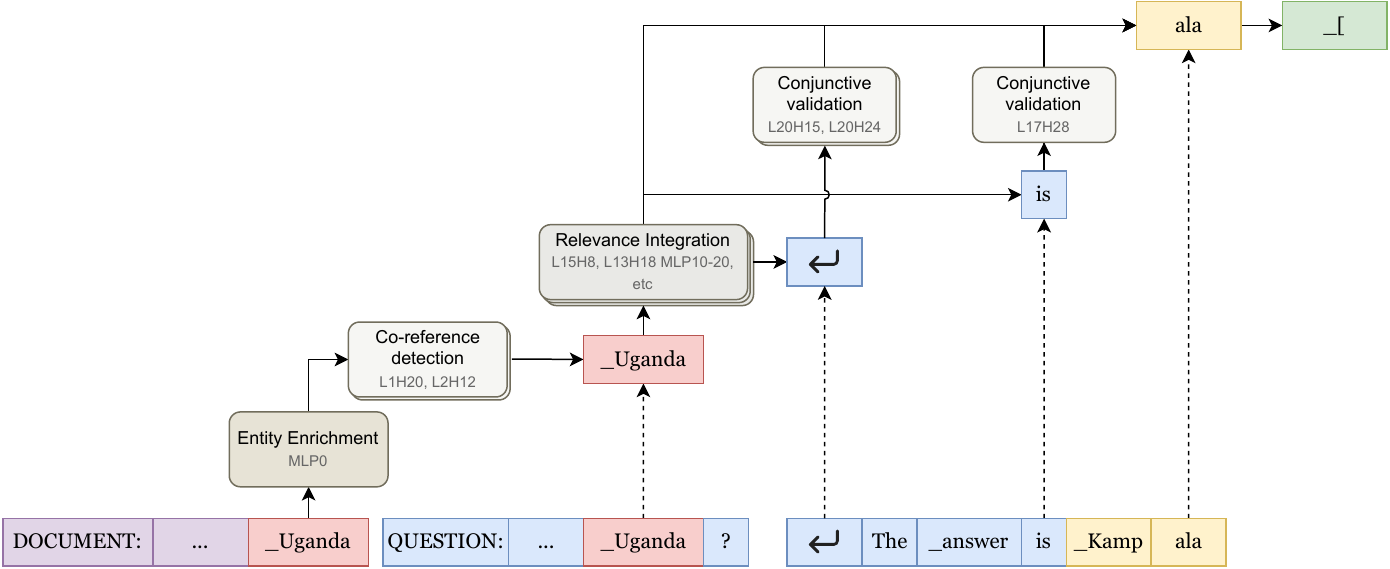}    \caption{Visualization of mechanistic processes that contribute to the citation generation.}
    \label{fig:mech_summary}
\end{figure}

\paragraph{Mechanistic Summary.} Inline citation emerges from: (1) early entity enrichment and duplication marking; (2) propagation through a mid-layer processing hub; (3) late aggregation that crosses a decision threshold at the answer token (see Figure~\ref{fig:mech_summary}). Multiple conjunctive (AND-gated) checks produce fragility: answer-token correctness is necessary but insufficient unless prior relevance propagation is intact. The ensemble is asymmetric: easy to disrupt, hard to induce.

\subsection{Steering the Mechanism: Targeted Interventions}
\begin{figure}[t]
	\centering
	\begin{subfigure}[b]{0.48\textwidth}
		\centering
		\includegraphics[width=\linewidth]{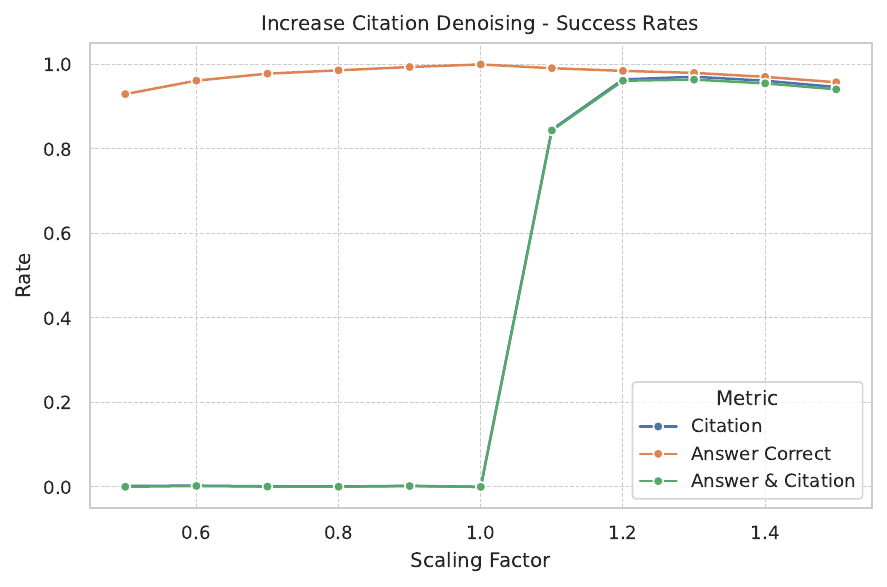}
		\caption{Inducing missed citations (threshold 0.10).}
	\end{subfigure}\hfill
	\begin{subfigure}[b]{0.48\textwidth}
		\centering
		\includegraphics[width=\linewidth]{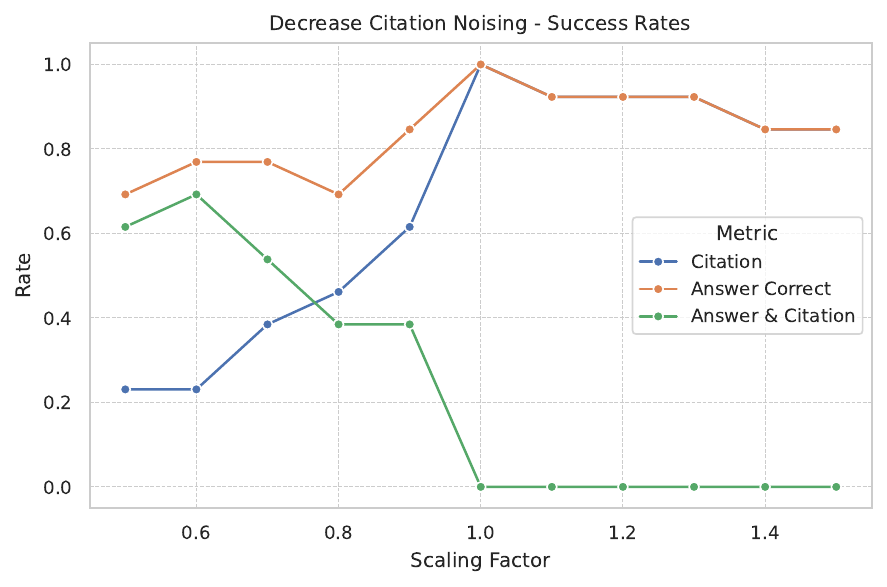}
		\caption{Suppressing spurious citations (threshold 0.20).}
	\end{subfigure}
	\caption{Targeted PopQA repairs via scaling identified components.}
	\label{fig:popqa_repairs}
\end{figure}

Having localized the ensemble, we tested whether scaling (multiplicative re-weighting) the outputs of identified components can correct two failure modes: \emph{missed citations} and \emph{spurious citations} in PopQA-derived contexts (Figure~\ref{fig:popqa_repairs}).

\paragraph{Repairing Missed Citations (PopQA).} Amplifying pro-citation (denoising-identified) components reliably induced missing citations (left panel Figure~\ref{fig:popqa_repairs}). Using a moderate importance threshold (logit difference > 0.10) and scaling factor \(\alpha = 1.2\), citation recovery exceeded 90\% with negligible impact on answer token correctness. A stricter component set (logit difference > 0.20) required larger scaling to approach similar gains, reflecting reduced redundancy.

\paragraph{Suppressing Spurious Citations (PopQA).} Down-scaling necessary (noising-identified) components effectively removed incorrect citations (right panel Figure~\ref{fig:popqa_repairs}). At threshold 0.20 with \(\alpha = 0.6\), 69\% (9/13) of spurious citations were suppressed, with only a minor decline in factual correctness. These complementary interventions confirm controllability.

\subsection{Scaling Across the Full PopQA Distribution}
\begin{figure}[t]
    \centering
    \begin{subfigure}[b]{0.53\textwidth}
        \centering
        \includegraphics[width=\linewidth]{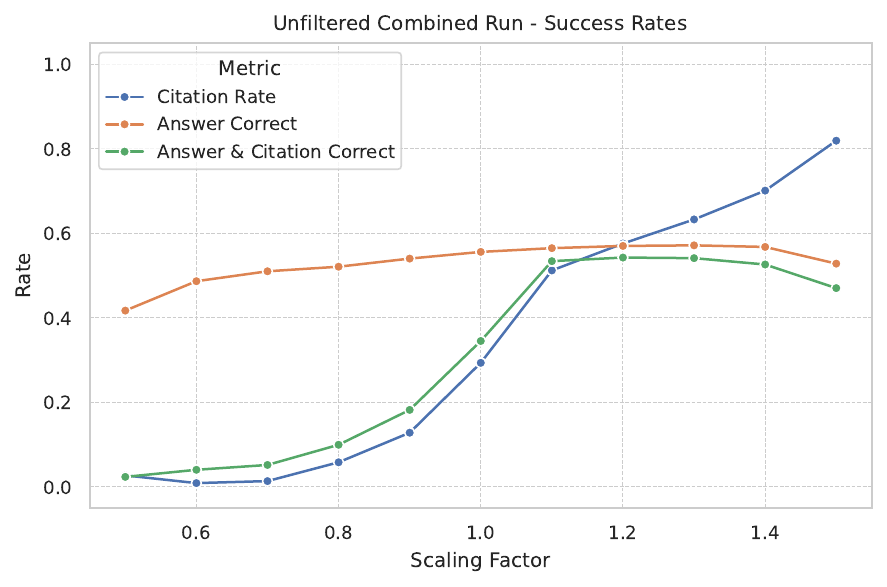}
        \caption{Distribution-level PopQA intervention: combined success metric (answer correct + cite only when relevant). Peak at modest amplification (\(\alpha=1.1\)).}
        \label{fig:popqa_combined}
    \end{subfigure}\hfill
    \begin{subfigure}[b]{0.44\textwidth}
        \centering
        \includegraphics[width=\linewidth]{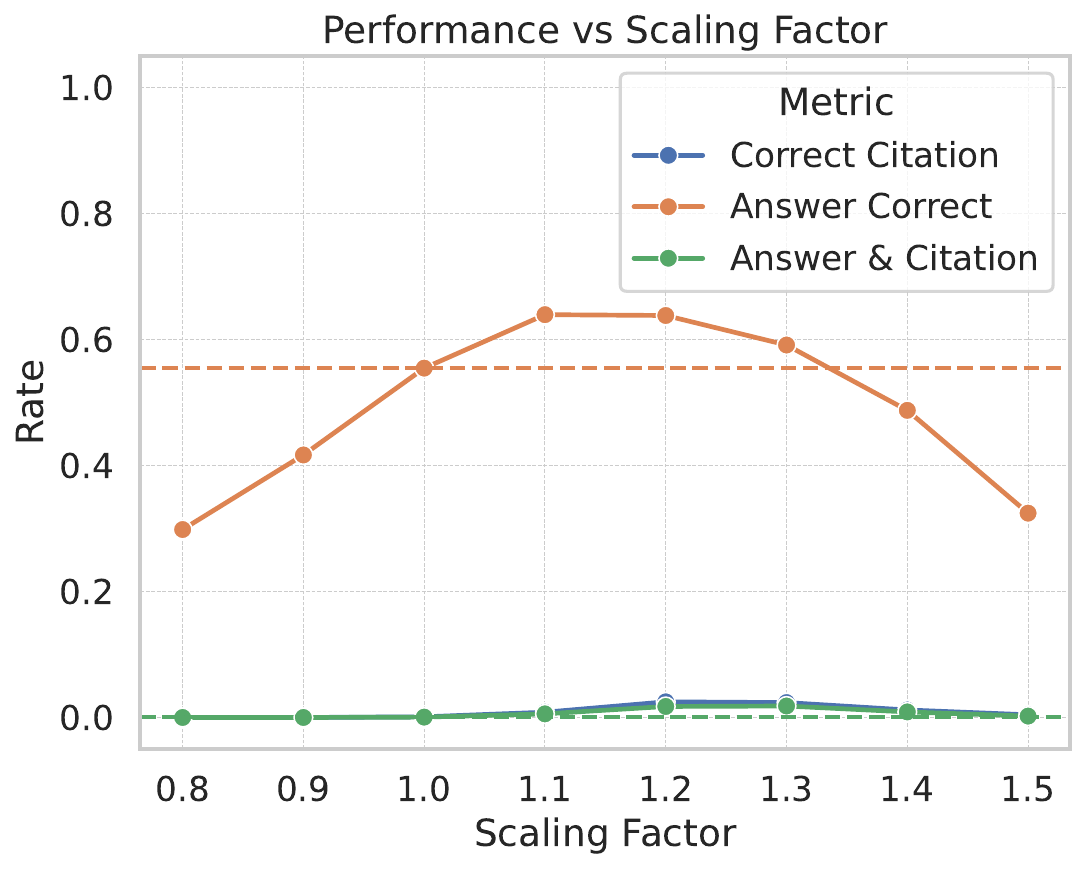}
        \caption{HotpotQA generalization: scaling identified components increases strict correct citation and joint correctness despite low absolute ceiling.}
        \label{fig:hotpotqa_scaling}
    \end{subfigure}
    \caption{Generalization of identified components from PopQA to HotpotQA. Left: PopQA distribution-level effect; right: multi-document HotpotQA transfer.}
    \label{fig:popqa_hotpotqa_combined}
\end{figure}

To test whether the identified ensemble generalizes beyond our controlled experimental setup to the broader dataset, we applied a unified intervention (combined important components, threshold 0.10) across the full PopQA set to test distribution-level shifts. Scaling upward (\(\alpha > 1.0\)) increased correct citation rates in relevant-document runs while slightly elevating spurious citations in irrelevant-document runs; scaling downward reduced both. A shallow optimum emerged at \(\alpha = 1.1\), maximizing a combined metric (Figure~\ref{fig:popqa_combined}). This indicates the ensemble encodes a global, not merely instance-specific, control dimension over citation policy.

% \begin{figure}[t]
% 	\centering
% 	\includegraphics[width=0.5\linewidth]{images/results/evaluation/popqa/unfiltered/unfiltered_combined_run_success_rates.pdf}
% 	\caption{Distribution-level PopQA intervention: combined success metric (answer correct + cite only when relevant). Peak at modest amplification (\(\alpha=1.1\)).}
% 	\label{fig:popqa_combined}
% \end{figure}

\subsection{Generalization to Multi-Document Reasoning (HotpotQA)}
We evaluated the same component set (no retuning) on 6,808 HotpotQA examples (multi-hop, multi-document). Baseline strict ``correct citation'' rate (all and only required documents) was \(0.001\). Amplifying with \(\alpha = 1.2\) increased this more than 20-fold to \(0.024\) (Figure~\ref{fig:hotpotqa_scaling}). The joint (answer+correct citation) rate improved from \(0.00059\) (4/6808) to \(0.01821\) (124/6808) at \(\alpha = 1.3\), a 31× relative gain though still low in absolute terms. Answer-string correctness also rose (0.55 → 0.64), and partial successes (some but not all required documents) increased while over-citation remained limited. Formatting near-misses (e.g., ``from Document 1'') suggest undercounting.

% \begin{figure}[t]
% 	\centering
% 	\includegraphics[width=0.5\linewidth]{images/results/evaluation/hotpotqa/hotpotqa_intervention_rates_0.1.pdf}
% 	\caption{HotpotQA generalization: scaling identified components increases strict correct citation and joint correctness despite low absolute ceiling.}
% 	\label{fig:hotpotqa_scaling}
% \end{figure}

\subsection{Implications}
\paragraph{Apparent Reasoning vs. Internal Mechanism.} A key finding is the potential disconnect between the model's output, which may appear to be the result of careful reasoning, and its internal computational pathway. An observer seeing a factually correct answer followed by a citation might assume a logical sequence: the model found the fact in the source, generated the answer from it, and therefore cited it. Our work mechanistically demonstrates that this is not necessarily the case. The citation can be generated by a separate, heuristic-based process that runs in parallel to answer generation. This suggests that LLM outputs can create a misleading narrative of their own reasoning, a critical insight for assessing the reliability of AI systems. A model can generate the \textit{appearance} of faithfulness while its internal mechanics follow a different logic, a phenomenon also observed in previous work on chain-of-thought reasoning~\citep{chen_reasoning_2025, turpin_language_2023}. Auditing the final output is not enough; we must audit the computational pathway that produced it.

\paragraph{RAG and Trustworthy Attribution.} The promise of RAG systems, particularly in high-stakes domains, hinges on their ability to provide verifiably grounded answers. Our investigation into the mechanics of inline citation reveals insights into the trustworthiness of this attribution method by examining the model's internal mechanisms against a strict definition of faithfulness.

A truly faithful citation would be mechanistically dependent on the model using the document to construct the answer. We would therefore hypothesize that the model's representation of the answer itself, at the moment of generation ($T_{ans}$), would be a critical nexus in the citation circuit. Our findings challenge this hypothesis. Activation patching revealed that while many components contribute to the final decision, the representation at the $T_{ans}$ token was not one of the most decisive drivers. This suggests the model's decision to cite may be driven by a sophisticated heuristic rather than being a direct result of the answer-generation process itself. This raises significant questions about whether prompted citations in their current form can truly deliver the level of verifiable grounding that RAG systems promise.

\subsection{Theory of Change}

RAG systems promise unprecedented capabilities in making complex information more accessible, through features like simplified language and conversational explanations. 
Their use in applications of societal decision-making processes, ranging from healthcare~\cite{yang2023large} and legal systems~\cite{seabrooke2024survey} to education~\cite{kasneci2023chatgpt}, seemingly enables people that might traditionally have had little access to specialized information, to inform themselves on these topics.

Yet, for AI systems to truly serve all of society, not just privileged groups with technical resources, we must ensure they are trustworthy and contain accurate information. Citations promise a way to decide whether or not to trust the generated information with the possibility to verify it. Yet, unfaithful citations may disproportionately impact communities that rely on AI systems as primary information sources but lack the resources or expertise to independently verify citations. This dynamic could deepen existing information disparities.
% The implications of unfaithful citations extend into crucial matters of algorithmic accountability and social impact. %Our findings raise important questions about how the reliability of explanations affects user trust in AI systems. 
Moreover, misleading citations can actually enhance user trust in incorrect AI responses, as explanations, even unfaithful ones, tend to increase perceived credibility \cite{sadeghi2024explaining}, creating a concerning form of AI-enabled misinformation that may be particularly difficult to detect and correct.
% When users encounter seemingly authoritative citations that are actually unfaithful, this creates a concerning form of AI-enabled misinformation that may be particularly difficult to detect and correct~\cite{sadeghi2024explaining}. %This risk is especially pronounced in high-stakes domains like healthcare, legal research, and policy-making, where citation accuracy is crucial for informed decision-making and accountability.

With this work we aim to take one step towards understanding and evaluating the citation generation in large language model, which we argue is essential for building trustworthy RAG systems. 
We see this work as a first stepping stone to an understanding of the internal process that leads to generating citations, which we hope will eventually lead to an evaluation framework that helps to identify trustworthy citation behavior.

Future work will need to investigate how mechanistic insights into the model predictions can be used to evaluate citation faithfulness in practice. Critics of mechanistic interpretability note that beyond scientific curiosity, not much practical impact has been derived from insights in the field.%Maria: There should be a reference here but I can not find a (pre-)print, just blog posts and twitter rants and I don't want to cite that. 
Another common criticism of explainable AI, which holds for our work as well is that techniques and findings are often built by, as well as, for engineers, without much consideration for the end user~\cite{miller_asylum}. 
For our findings to have real world impact, next steps should investigate how the knowledge about the model components that are responsible for generating citations can be used to infer citation faithfulness. 

Another limitation of our work is the open question of generalizability of our findings to other datasets, models and scenarios. Only if the insights of our work generalize, will it be possible to infer a reliable evaluation framework in the ever changing landscape of models and applications. 

Lastly we want to highlight the possible negative externalities of misuse of our research in in an adversarial setting. The knowledge about what input triggers a citation or which model components need to be activated to generate a citation, might be used to insert unfaithful citations that promote certain statements. This misuse of gained insight into model behavior unfortunately is an intrinsic downside of any interpretability research~\cite{winninger_using_2025}, and we hope that advantages will outweight those risks.

\section{Conclusion}\label{chapter:conclusion}

In this study, we used mechanistic interpretability to reverse-engineer how Llama-3.1-8B-Instruct decides whether to emit inline citations in a RAG context. Through systematic activation patching, we identified internal components which causally drive citation behavior and tested whether manipulating them could correct citation failures.
We found that citation generation is implemented by a distributed ``attributional ensemble'' of attention heads and MLPs that operates through conjunctive checks rather than a single specialized module. This ensemble relies heavily on shallow heuristics, particularly entity co-reference matching between document and question, rather than deep semantic integration during answer construction.

We argue that this suggests that inline citations can create an illusion of verifiability even when answers are not genuinely grounded in sources. % For communities that rely on RAG systems as primary information sources but lack resources to independently verify citations, this disconnect between apparent reasoning and internal mechanism poses a serious equity concern. 
This work can be seen as a first step towards making these failure modes mechanistically legible and toward an evaluation frameworks that can distinguish faithful from unfaithful attribution.
% Making these failure modes mechanistically legible is a first step toward evaluation frameworks that can distinguish faithful from unfaithful attribution essential for deploying AI systems that are trustworthy for all users, not just those with technical expertise.

Our study has important limitations: we analyzed a single 8B model in a controlled, single-document setting; activation patching localizes components but not directed information pathways; and we did not directly map the answer-generation circuit to test mechanistic overlap with citation generation.

Future work should refine the ensemble into a complete directed circuit via path patching, explicitly map and compare the answer-generation and citation circuits to operationalize mechanistic faithfulness, and replicate across model architectures, scales, and RAG-finetuned systems to assess whether these mechanisms are universal or model-specific.

\begin{credits}
% \subsubsection{\ackname} A bold run-in heading in small font size at the end of the paper is
% used for general acknowledgments, for example: This study was funded

\subsubsection{\discintname}
The authors have no competing interests to declare that are
relevant to the content of this article.
\end{credits}




% LNCS Homepage, \url{http://www.springer.com/lncs}, last accessed 2023/10/25
% \end{thebibliography}

\bibliographystyle{splncs04}

\bibliography{bibentries}

\end{document}